\documentclass[12pt]{iopart}

\usepackage{iopams}
\expandafter\let\csname equation*\endcsname\relax
\expandafter\let\csname endequation*\endcsname\relax

\usepackage{amsmath}
\usepackage{natbib}
\usepackage{graphicx, float}
\usepackage{amsmath}
\usepackage{caption}
\usepackage{hyperref}
\usepackage{wrapfig}
\usepackage{xcolor}
\usepackage[shortlabels]{enumitem}
\usepackage{verbatim} 

\usepackage{soul} 

\newcommand{\Real}{\mathbb{R}}

\newtheorem{theorem}{Theorem}[section]

\def\spacingset#1{\renewcommand{\baselinestretch}%
{#1}\small\normalsize} \spacingset{1}

\begin{document}

\title[]{A Statistical Primer on Classical Period-Finding Techniques in Astronomy}

\author{Naomi Giertych\textsuperscript{\dag}, Ahmed Shaban\textsuperscript{\ddag}, Pragya Haravu\textsuperscript{\dag}, Jonathan P Williams\textsuperscript{\dag}}

\address{\textsuperscript{\dag}Department of Statistics, North Carolina State University, Raleigh, NC, USA}
\address{\textsuperscript{\ddag}Department of Physics, North Carolina State University, Raleigh, NC, USA}
\ead{ngierty@ncsu.edu}
\vspace{10pt}
\begin{indented}
\item[]
\end{indented}

\begin{abstract}
The aim of our paper is to investigate the properties of the classical phase-dispersion minimization (PDM), analysis of variance (AOV), string-length (SL), and Lomb-Scargle (LS) power statistics from a statistician's perspective. We confirm that when the data are perturbations of a constant function, i.e. under the null hypothesis of no period in the data, a scaled version of the PDM statistic follows a beta distribution, the AOV statistic follows an F distribution, and the LS power follows a chi-squared distribution with two degrees of freedom. However, the SL statistic does not have a closed-form distribution. We further verify these theoretical distributions through simulations and demonstrate that the extreme values of these statistics (over a range of trial periods), often used for period estimation and determination of the false alarm probability (FAP), follow different distributions than those derived for a single period. We emphasize that multiple-testing considerations are needed to correctly derive FAP bounds. Though, in fact, multiple-testing controls are built into the FAP bound for these extreme-value statistics, e.g. the FAP bound derived specifically for the maximum LS power statistic over a range of trial periods. Additionally, we find that all of these methods are robust to heteroscedastic noise aimed to mimic the degradation or miscalibration of an instrument over time.  Finally, we examine the ability of these statistics to detect a non-constant periodic function via simulating data that mimics a well-detached binary system, and we find that the AOV statistic has the most power to detect the correct period, which agrees with what has been observed in practice.
\end{abstract}

\vfill

\section{Introduction}

Periodic variable objects, i.e. objects whose properties change over regular intervals of time, are ubiquitous in the universe and are a source of continuous study for astronomers. For example, the periodic variation of pulsating stars allow astronomers to calculate the distance to these stars and use them as standard candles or for cosmology, like measuring the Hubble constant \citep{Bahcall2015hubble, Riess1998AcceleratingUniverse}. Alternatively, periodic changes in a star's brightness could be used to classify a star as being part of a binary system, e.g. \cite{xu2022new}, or indicate the presence of a planetary companion, e.g. \cite{rappaport2012possible, Lafarga2020}. Outside of stars, the periodic rotation of Cold Classical Kuiper Belt Objects contain information about planet formation and the development of our solar system during the first 100 million years after the Sun’s birth \citep{thirouin2019light}. Thus, accurately characterizing a periodic signal in observed data is of vital importance to astronomers. 

Unfortunately, time series analysis techniques historically employed for signal processing only apply to data observed at equally-spaced time intervals, but astronomers often have data where the observations are unevenly spaced in time, e.g. \cite{capistrant2022population, Feigelson2021, mowlavi2022gaia, Suveges2014}. This constraint forced astronomers to develop their own methods for period detection such as the AOV \citep{Schwarzenberg1989}, PDM \citep{Stellingwerf1978}, SL \citep{Dworetsky1983}, and the LS power \citep{Lomb1976, Scargle1982} statistics.

However, because the field of statistics diverged from astronomy in the late 19th century, statisticians had little involvement in the development of these methods. This lack of input from statisticians has led to concerns over the statistical properties of these classical astrostatisical methods for period detection. For instance, in reference to the PDM statistic, \cite{Feigelson2021} state:
\begin{quote}
``None of these statistical issues [e.g. non-Gaussian, correlated error and multiple-testing] have been examined by mathematical statisticians despite the use of the PDM in $\sim$1,500 astronomical papers over four decades.''
\end{quote} Though, of course, these concerns are also valid for the other methods introduced above. Our paper begins to fill this gap by investigating and comparing the theoretical and empirical properties of these classical statistics and aims to open a dialogue within the astronomical community about the appropriate use of them. To the best of our knowledge, no such discussion and comparison has been published from a statistician's perspective, though some have been published by astronomers, e.g. \cite{hara2023statistical, Heck1985, VanderPlas2018}. 

The organization of the paper is as follows. Section \ref{sec:astro} briefly reviews some examples of where the discussed techniques are commonly employed in astronomy. Section \ref{sec:model} discusses the statistical model of the raw data. Section \ref{sec:ht} discusses the hypothesis test of detecting a periodic signal from data. Section \ref{sec:pb_methods} details the phase-based methods; specifically the AOV, PDM, and SL methods are discussed. Section \ref{sec:npb_methods} details the LS power. Section \ref{sec:sims} empirically demonstrates through simulation studies the theoretical properties discussed in Sections \ref{sec:pb_methods} and \ref{sec:npb_methods}, explores the robustness of these methods to challenges faced in real-world datasets, and briefly discusses the power of these statistics to detect periodic signals encountered by astronomers. Finally, Section \ref{sec:conc} finishes with some concluding remarks. Code and source files to reproduce all figures are publicly available at \href{https://ngierty.github.io/research.html}{https://ngierty.github.io/research.html.}

\section{Astronomical Context}\label{sec:astro}

In this section, we briefly demonstrate where the statistical methods for period detection discussed in this paper are commonly employed in astronomy. In particular, we highlight that these methods have been used for determining the periods of variable stars and other objects inside and outside of our solar system.

Variable stars are divided into two groups based on their type of variability: 1) extrinsic variability and 2) intrinsic variability. The extrinsic variability in a star's brightness is due to some external change like rotation of the star or eclipses. For example, a star's brightness changes as a result of a star's rotation because dark solar spots on the surface of the star appear and disappear from view \citep{griffiths2018rotatingVariables}. Alternatively, in an eclipse, the light we see from a star is blocked by another object, often a companion planet or star, and appears as a dip in the light curve \citep{Tohline2002BinaryStarsOrigin}. One of the most common types of these variable stars are eclipsing binaries, where there are two stars orbiting each other. Multiple studies have determined the periods of such systems using the methods studied here in the Milky Way and beyond, e.g. \cite{Stassun2004EclipsingBinary, Graczyk2011OGLEIIIBinariesLMC, Prsa2011KeplerEclipsingI} and \cite{xu2022new}. Additionally, these methods have occasionally been used to detect the presence of extra-solar planets. For instance, \cite{AngladaEscude2016TerrestialExoplanet} discovered a terrestrial exoplanet candidate around \textit{Proxima Centauri} using the LS power.

On the other hand, intrinsic variability is from variations in the physical properties of the star. These variables are classified into pulsating, eruptive, or cataclysmic. A pulsating variable star's radius contracts and expands periodically which changes the star's brightness \citep{Eyer2008VariableStars}. RR Lyrae and Cepheids are examples of pulsating stars, and many studies used AOV, PDM, and LS to estimate their period, study the period's relation to the star's luminosity, and estimate the star's distance from Earth \citep{Matsunaga2006CepheidsClusters, Nemec2013RRLyrae, Udalski1999CepheidsLMC, Udalski2018GalacticCepheids, Torrealba2015RRLyrae, Gerke2011CepheidsM81, Shappee2011CepheidDistanceM101}. An eruptive variable star's surface goes through eruptions and accretions of matter, causing variations in brightness \citep{Good2003eruptive}. Protostars, Orion variables, and gas giants are examples of objects that are frequently classified as eruptive variables. Finally, a cataclysmic variable star goes through a sudden change in the surface brightness of a star at the end of main-sequence life \citep{Knigge2011CataclysmicEvolution}. Novae and supernovae are examples of this type of variation \citep{Baade1934Supernovae, Chomiuk2021ClassicalNovae}. While less frequent, the statistical methods discussed in this paper have also been used as a first step to identifying stars in the eruptive and cataclysmic variable star sub-categories, e.g. \cite{alfonso2012first}.

Finally, the methods examined here have also been used to determine the periodic properties of non-stellar objects. For example, several studies studied the periodicity of objects within our solar system such as the quasi-periodic variation in solar wind speed and geomagnetic activity \citep{Mursula2000SolarWindSpeedVariation}, the periodicity of outburst activity in comets \citep{TrigoRodriguez2010Comet}, and the rotation period of Kuiper Belt Objects \citep{Lacerda2008KupierBeltObject}. Additionally, these methods have been used to determine the periodicity of more exotic astrophysical systems such as Fast Radio Bursts (FRBs) \citep{Rajwade2020FRB}, short-period variable quasars \citep{Charisi2016VariableQSO}, $\gamma$-ray bright Blazers \citep{Bhatta2020Blazar, Bhatta2021Blazars, OteroSantos2023Blazars}, methanol masers \citep{Goedhart2014Masers}, and $\gamma$-ray periodicity in active galactic nuclei (AGN) \citep{Penil2020PeriodicityAGN}.

This brief summary shows the importance and diverse use of the statistical methods reviewed in this paper for studying the time series observations of many astrophysical objects and phenomena.


\section{Statistical Model of Raw Data}\label{sec:model}

Period detection methods can be used for any numerical observation that a researcher believes comes from a periodic signal (including a constant signal) that has been observed with some noise. For example, to determine whether a star of interest is a variable star, astronomers measure the star's {\em flux} or {\em apparent magnitude}, both standardized measurements of the amount of light Earth receives from a star, for several time points. This flux or magnitude is often assumed to come from a periodic signal but observed with additive noise. This section details how such an observed value can be represented mathematically in the time domain or the frequency domain, which are then used to construct the PDM and AOV statistics in Section \ref{sec:pb_methods} and the LS power in Section \ref{sec:npb_methods}.

Specifically, the statistical model for a numerical observation, $X_t$, at time $t \in \{t_1, \dots, t_T\}$ is
\begin{equation}\label{eq:periodx}
X_t = h(t) + \epsilon_t,
\end{equation} 
where $h(t)$ is a periodic function oscillating around zero with true period $p_0$ (i.e. $h(t) = h(t + kp_0)$, for any integer $k$), and $\epsilon_{t} \overset{\text{iid}}{\sim} N(0, \sigma^2)$ for some $\sigma > 0$; in other words, we assume the random noise is a realization of an independent and identically distributed Gaussian random variable. This periodic function can be written in either the time domain, as presented in equation (\ref{eq:periodx}), or can be transformed into the frequency domain using the discrete Fourier transform defined as 
\begin{equation}\label{eq:freq_domain}
\hat{X}(f) := \sum_{t = t_1}^{t_T} X_t e^{-2\pi i f t/T} = \sum_{t = t_1}^{t_T} X_t [\cos(2\pi f t/T) -i \sin(2\pi f t/T)]
\end{equation}
for frequency $f \in \Real$, imaginary $i$, and equally spaced time points $t$.

There are two important aspects to note about the data generating model in equation (\ref{eq:periodx}). First, a constant function is considered a periodic function with $p_0 = \infty$ \citep{Priestley1981}. While this is not of direct interest to astronomers, it is important in calculating the FAP, often referred to as the p-value in statistics literature. This issue is discussed more below in Section \ref{sec:ht} and Section \ref{sec:pb_methods}. Second, assuming $\epsilon_{t} \overset{\text{iid}}{\sim} N(0, \sigma^2)$ is a fairly strong assumption which is not always met in practice \citep{Feigelson2021}. Specifically, the assumption of homoscedastic variance is often violated in astronomical data sets due to measurement error; fortunately, however, astronomers can typically calculate their measurement error. This information can be used to rescale the data to correct for heteroscedastic measurement errors. In particular, given heteroscedastic data generated according to model (\ref{eq:periodx}) with $\epsilon_t \overset{\text{indep}}{\sim} N(0, \sigma_t^2)$, the transformed data $\Tilde{X}_t = \sigma_t^{-1} X_t$ has homoscedastic variance. However, care must be taken as correcting for heteroscedasticity in this way assumes that all noise in the process has been accounted for in determining $\sigma_t$. 

Finally, if either the homoscedasticity or the Gaussianity assumption is violated, the calculated FAPs will misrepresent the actual false positive rate. This could lead the practitioner to a belief about a spurious result or to overlook a meaningful result. There are statistical tests that can and should be used to test the assumptions of Gaussianity, e.g. D'Agostino's $K$-squared test \citep{d1970transformation} and the Shapiro-Wilk test \citep{shapiro1965analysis}, and homoscedasticity, e.g the Breusch-Pagan test \citep{breusch1979simple} or the Goldfeld–Quandt test \citep{goldfeld1972nonlinear}. However, these tests often have additional assumptions and may not be appropriate for small sample sizes. Unless the practitioner has prior knowledge of the validity of these assumptions, they should avoid computing or interpreting analytical FAPs (those based on the parametric distributions discussed below) and instead use FAPs calculated from bootstrapping the data. Though, the practitioner should be warned that bootstrapping is not a panacea, even if the computational burden is low. First, the variability in the bootstrapped estimates is smaller than the variability that would be expected in the estimates obtained from new samples. This is because the population from which the bootstrapped samples are drawn is fixed as the observed sample. While this issue is not particularly problematic for medium and large sample sizes, practitioners should be wary of applying this method for smaller sample sizes; see \cite{chernick2011bootstrap} for recommendations. Second, the bootstrap method may fail to provide accurate estimates under certain transformations of the data, e.g. the minimum or maximum of a set \citep{chernick2011bootstrap}. However, there have been proposed remedies to the bootstrapping procedure to address such limitations and maintain desirable properties of the bootstrap under certain conditions \citep{chernick2011bootstrap, fukuchi1994bootstrapping}.

\section{Hypothesis Test for Periodicity}\label{sec:ht}

To perform a test for periodicity, one must test whether the data come from a non-constant periodic function, i.e. test if $p_0 < \infty$, e.g. \cite{Dworetsky1983, Schwarzenberg1989, Schwarzenberg1997} and \cite{Stellingwerf1978}. This can be written as a formal hypothesis test as follows:
\begin{equation} \label{eq:ht}
    H_0: p_0 = \infty \hspace{24pt} H_a: p_0 < \infty.
\end{equation} The FAP is then calculated using the distribution of the test statistic (e.g. PDM, AOV, etc.) assuming $H_0$ is true. As will be shown in Sections \ref{sec:pb_methods} and \ref{sec:npb_methods}, only the PDM, AOV, and LS power statistics have known closed-form distributions under $H_0$. Thus, for the remainder of the paper, when we discuss a test for periodicity, as well as calculating an FAP, we are doing so under (\ref{eq:ht}).

Finally, since the true period is unknown, many trial periods are tested. The FAP associated with the period that gives the minimum PDM or SL statistic or the maximum AOV statistic or LS power does not follow the same calculation as the FAP for an arbitrary period. Further, the FAP associated with the period that gives the extreme statistic, e.g. the maximum AOV, will automatically control for considering a range of potential periods and calculating statistics at each of them. This is discussed in more detail below. Issues of multiple-testing occur, for instance, if one period detection method is used to narrow a large range of periods to a more manageable set of trial periods, and then another period detection method is used only on the smaller set of trial periods to determine which periods are significant; or only the smaller set of periods is used in constructing downstream models which undergo their own model selection procedure, e.g. \cite{mowlavi2022gaia} and \cite{ritchie2022vlt}. 

\section{Phase-Based Methods} \label{sec:pb_methods}

\begin{figure}[h]
\centering
\includegraphics[scale=0.24]{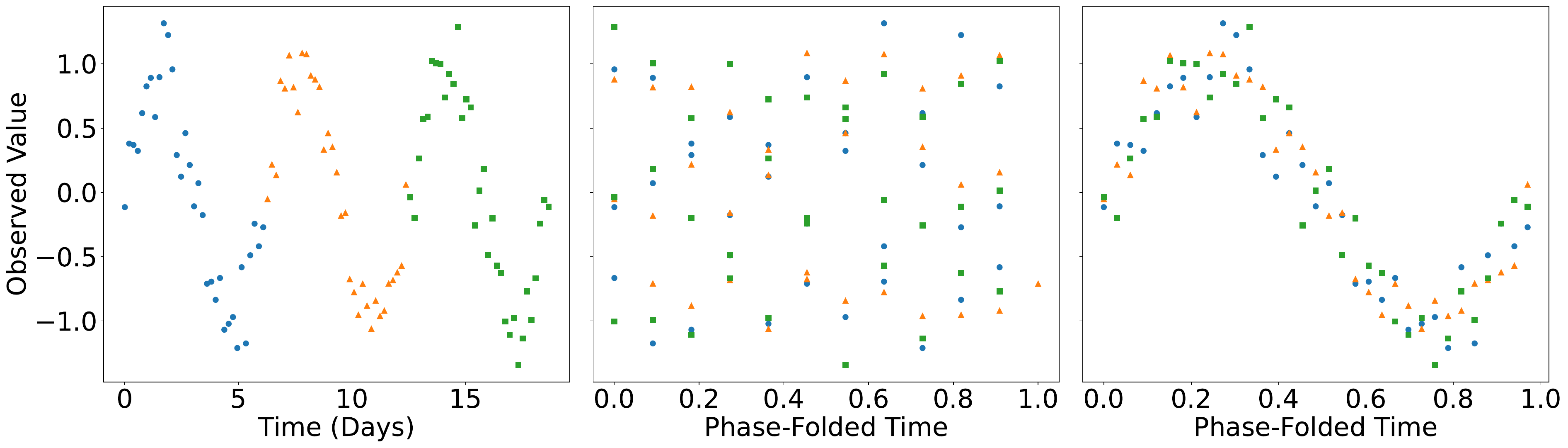}
\caption{\footnotesize This figure demonstrates how phase-folding can be used to construct tests of trial periods given some observations that come from a periodic function. The \textit{left panel} displays simulated observations generated as perturbations from a periodic function. The \textit{middle} and \textit{right panels} illustrate the result of phase-folding the observations according to an incorrect and the correct trial period, respectively. Clearly, if one were to bin the observations based on the phase-folded time then the bin variances would be smallest and the differences in bin means would be largest in the right-most figure. Similarly, a line connecting all observations in phase-based order would be shortest in the \textit{right panel}.}
\label{fig:phase_folding}
\end{figure}

Phase-based methods aim to determine the period of a periodic function based on noisy observations, and they leverage the idea that observations can be {\em folded} (or {\em phase-folded}) in time modulo the true period of the function. We will denote the phase-folded time as $\phi_t = (t \text{ mod } p)/p$; thus, the phase-folded-time--observation pair will be denoted $(\phi_t, X_t)$. Given a trial period, the observed data can always be phase-folded, but if the trial period is not close to the true period then the phase-folded values likely will not align with the function $h$ in phase. For example, the left panel of Figure \ref{fig:phase_folding} shows a set of arbitrary observations, e.g. magnitude or flux, that has been centered and scaled with their corresponding (arbitrary) observation times where centering and scaling, here, refers to subtracting the sample mean from each observation and then dividing by the sample standard deviation. The middle and right panels then show the same data phase-folded for an incorrect and correct trial period, respectively. Clearly, the data in the right panel aligns with the underlying function.

In the classical literature on these methods, \cite{Lafler1965, Stellingwerf1978, Schwarzenberg1989, whittaker1926calculus} suggest binning the phase-folded data into groups with similar phase-values and deriving statistics based on ratios of sums of squared deviations to test various trial periods. Specifically, the AOV statistic aims to measure the difference in the bin means and the PDM statistic aims to measure the differences between the observations within a bin. Alternatively, \cite{Burke1970} suggested measuring the fit of the trial period $p$ based on the Euclidean distance between neighboring time-observation pairs after phase-folding; this method is often referred to as the SL method. Examining Figure \ref{fig:phase_folding} again, clearly, the phase-folded observations in the right panel will have bin means that are the most different (i.e. the largest AOV), bin variances that are the smallest (i.e. the smallest PDM), and the smallest SL compared to other trial periods. We discuss these methods separately below.

\subsection{Sum-of-Squares Methods}

The basic methodologies for detecting non-sinusoidal periodic signals as developed in \cite{Stellingwerf1978} and \cite{Schwarzenberg1989} rely on a simple sum-of-squares decomposition. They are intuitively elegant, and the theory behind these methods, namely Cochran’s Theorem, is well-rooted in classical statistical results dating back to the early twentieth century \citep{monahan}. The purpose of this section is to provide a detailed account of the statistical theoretical foundation for these tests.

Namely, for a trial period $p$, the null and alternative hypotheses suggested by \cite{Stellingwerf1978} are formulated as
\begin{equation}\label{hypotheses}
\begin{split} 
H_0 &: \mu_1 = \cdots = \mu_r  \hspace{.25in} \\ 
H_a &: \text{at least one $\mu_i$ is different from the others} \hspace{.25in} \\
\end{split}
\end{equation} 
where $\mu_{1}, \dots, \mu_{r}$ are defined as the bin means for the phase-folded data modulo $p$. 
The bin means being equal (as in $H_0$) suggests no periodicity in the observed values, and so statistical evidence supporting the assertion of a periodic signal in the data is established if a test rejects $H_0$ in favor of $H_a$. Note, this is equivalent to the hypothesis test setup in equation (\ref{eq:ht}) after the data has been phase-folded and binned. The classical PDM and AOV test statistics are derived next. \footnote{Note, since the PDM statistic is a generalization of the Lafler and Kinman technique and a modification of the Whittaker and Robinson statistic \citep{Stellingwerf1978}, we will not discuss these.}

Let $X_{ij}$ be an observation, e.g. observed flux of a star, which has already been binned in-phase based on some trial period $p$, where $i \in \{1, \dots, r\}$ is the bin index, $r$ is some specified number of bins, $j \in \{1, \dots n_i\}$ is the observation index for the $i$-th bin, and $n_i$ is the number of observations in bin $i$.  Let $n := \sum_i n_i$ be the total number of observed time instances, and take $S_0^2$, $S_1^2$, and $S_2^2$ to denote the total sample variation, the across-bins mean variation, and the within-bin variation, respectively. Then, 
\begin{equation}\label{pdm_aov} 
\begin{split}
(n-1)S_0^2 &= \sum_{i = 1}^{r} \sum_{j=1}^{n_{i}} (X_{ij} - \bar{X})^2 \\
& = \sum_{i = 1}^{r} n_{i}(\bar{X}_{i} - \bar{X})^2 +  \sum_{i=1}^{r} \sum_{j=1}^{n_{i}} (X_{ij} - \bar{X_{i}})^2
= (r-1)S_1^2 + (n-r)S_2^2.
\end{split}
\end{equation} 
The PDM statistic proposed by \cite{Stellingwerf1978} and the AOV statistic proposed by \cite{Schwarzenberg1989} are defined, respectively, as
\begin{align}
T_{\text{PDM}} &:= S_2^2/S_0^2 \label{PDMstat} \\ 
T_{\text{AOV}} &:= S_1^2/S_2^2, \label{AOVstat} 
\end{align}
and the trial period $p$ is tested by determining the FAP, which is the tail probability of the sampling distributions of these statistics under the null hypothesis $H_0$.

Under $H_0$ and model (\ref{eq:periodx}), \cite{Stellingwerf1978} incorrectly argued that the PDM statistic followed some parameterization of the F distribution. The inappropriateness of the F distribution for the PDM statistic is discussed in \cite{Heck1985} and referenced in \cite{Nemec1985} with inaccurate mathematical justification for why it does not follow an F distribution. These authors erroneously claim that the PDM statistic does not follow an F distribution because the numerator, $S_2^2$, does not follow a chi-squared distribution. They are misled in remarking that the matrix associated with its quadratic form is not idempotent.  Subsequently, however, \cite{Schwarzenberg1989} correctly concluded that the PDM statistic does {\em not} follow a F distribution since the numerator and denominator are not independent (as is clear from equation (\ref{pdm_aov})).  In a followup paper, \cite{Schwarzenberg1997} argues that a scaled version of the PDM statistic, specifically $\frac{n-r}{n-1}T_{\text{PDM}}$, follows the beta distribution with shape parameters $\frac{n-r}{2}$ and $\frac{r-1}{2}$, assuming $H_0$ is true.  Additionally, \cite{Schwarzenberg1989} accurately states the AOV statistic follows the F distribution with degrees of freedom $r-1$ and $n-r$ under $H_0$.

Arguing that $T_{\text{AOV}} \sim \text{F}_{r-1,n-r}$ is a direct application of Cochran's theorem which is stated below; the proof is omitted. The statement roughly follows the lecture notes of \cite{Feng2015} for clarity and accessibility to the reader; alternative formulations and proofs can be found in statistical texts such as \cite{monahan}.

\begin{theorem}(Cochran's theorem)
    Let $X_1, \dots X_n \overset{iid}{\sim} N(0, \sigma^2)$, and suppose that
    \begin{equation*}
        \sum_{i=1}^n X_i^2 = Q_1 + \cdots + Q_k
    \end{equation*} where $Q_1, \dots, Q_k$ are positive semi-definite quadratic forms in $X_1, \dots, X_n$. That is, 
    \begin{equation*}
        Q_i = X'A_i X, \hspace{12pt} \text{for $i \in \{1, \dots, k\}$}
    \end{equation*} where $X' = (X_1, \dots, X_n)$ and $A_i$ are positive semi-definite matrices with $r_i = \text{rank}(A_i)$. If $\sum_i r_i = n$, then
    \begin{enumerate}
        \item $Q_1, \dots, Q_k$ are independent, and
        \item $Q_i \sim \sigma^2 \chi_{r_i}^2$ for $i \in \{1, \dots, k\}$ where $\chi_{r_i}^2$ denotes the chi-squared distribution with $r_i$ degrees of freedom.
    \end{enumerate}
\end{theorem}

Note, equation (\ref{pdm_aov}) can be expressed in terms of the quadratic forms 
\begin{equation}\label{matrix_pdf_aov}
\underbrace{X'(I_n - n^{-1} J_n)X}_{(n-1)S_0^2} = \underbrace{X'(P_V - n^{-1} J_n)X}_{(r-1)S_1^2} + \underbrace{X'(I_n - P_V)X}_{(n-r)S_2^2},
\end{equation}
where $X := (X_{11}, \dots, X_{1n_1}, \dots, X_{r1}, \dots, X_{rn_r})'$, $I_n$ is the $n\times n$ identity matrix, $J_n$ is an $n \times n$ matrix of ones, and $P_V := V(V'V)^{-1}V'$ is the projection matrix onto the column space of $V := \text{diag}\{1_{n_1}, \dots, 1_{n_r}\}$. Note, $V$ is simply a matrix whose columns correspond to a bin number and rows are indicators of whether an observation belongs to a particular bin; in other words, $V_{i,j}$ (the $i$th row and $j$th column of $V$) is an indicator for whether observation $i$ has been phase-folded into bin $j$. If $p_0 < \infty$, we take $V$ as the matrix of indicators if the data are phase-folded modulo $p_0$. More simply, $V$ is the design matrix in the linear regression $X = V\mu + \epsilon$ with $\mu = (\mu_{1},\dots,\mu_{r})'$ and $\epsilon := (\epsilon_{11}, \dots, \epsilon_{1n_1}, \dots, \epsilon_{r1}, \dots, \epsilon_{rn_r})'$.

Finally, Cochran's theorem applies to establish that $S_1^2/\sigma^2 \sim \chi_{r-1}^2$ independently of $S_2^2/\sigma^2 \sim \chi_{n-r}^2$ because from equation (\ref{matrix_pdf_aov}),
\[
X'X = X'\underbrace{n^{-1}J_n}_{=: A_{1}}X + X'\underbrace{(P_V - n^{-1}J_n)}_{=: A_{2}}X + X'\underbrace{(I_n - P_V)}_{=: A_{3}}X,
\]
with $A_{1}, A_{2}$, and $A_{3}$ all idempotent, $A_1 + A_2 + A_3 = I_n$, and rank$(A_{1}) = 1$, rank$(A_{2}) = r-1$, and rank$(A_{3}) = n-r$.

Arguing that $T_{\text{PDM}} \sim \text{beta}\big(\frac{n-r}{2},\frac{r-1}{2}\big)$ follows by a series of arithmetic steps to obtain
\[
\begin{split}
F_{\frac{n-r}{n-1}T_{\text{PDM}}}(u) & = P\Big(\frac{n-r}{n-1}T_{\text{PDM}} \le u\Big) 
= 1 - F_{T_{\text{AOV}}}\Big(\frac{n-r}{r-1}\{u^{-1} - 1\}\Big). \\
\end{split}
\]
Then, since $T_{\text{AOV}} \sim \text{F}_{r-1,n-r}$, taking derivatives on both sides yields, for $u \in [0,1]$,
\[
f_{\frac{n-r}{n-1}T_{\text{PDM}}}(u) = \frac{  u^{\frac{n-r}{2} - 1}(1 - u)^{\frac{r-1}{2} - 1}  }{B(\frac{n-r}{2},\frac{r-1}{2})};
\]the density function of the $\text{beta}\big(\frac{n-r}{2},\frac{r-1}{2}\big)$.

Observe that the $\sigma^2$ terms cancel when constructing $T_{\text{AOV}}$ and $T_{\text{PDM}}$.
A point of caution is that this cancellation will not occur when the errors are heteroscedastic, invalidating the appropriateness of relying on critical values from the F or beta distribution, respectively. In the heteroscedastic setting, the practitioner would need to know the standard deviations to scale each observation as discussed in Section \ref{sec:model}.

Under $H_0$, the PDM and AOV statistics will be from the $\text{beta}\big(\frac{n-r}{2},\frac{r-1}{2}\big)$ and $\text{F}_{r-1,n-r}$ densities, respectively, regardless of the trial period. However, if $H_a$ is true, the distributions of $T_{\text{AOV}}$ and $T_{\text{PDM}}$ depend on whether the trial period $p$ is the true period $p_0$ or not. If $p = p_0$, it is established from well-known statistical theory that $T_{\text{AOV}}$ follows the non-central F distribution $\text{F}_{r-1,n-r}\big(\frac{1}{2\sigma^2}\mu'A_{2}\mu\big)$ and $T_{\text{PDM}}$ follows the Type II non-central beta distribution $\text{beta}_{\frac{n-r}{2}, \frac{r-1}{2}}\big(\frac{1}{2\sigma^2} \mu'A_{2}\mu\big)$, see \cite{monahan} and \cite{chattamvelli1995} for details.  In the case that $p \ne p_0$, the flux values have been folded modulo the wrong period, and so the true data generation equation has the form $X = W\mu + \epsilon$ for some design matrix $W \ne V$.  Accordingly, $T_{\text{AOV}}$ and $T_{\text{PDM}}$ follow doubly non-central F and doubly non-central beta distributions, respectively \citep{chattamvelli1995f}. While the distributions that arise under $H_a$ could theoretically be used to determine the power of the tests, assumptions would need to be made on $\mu$ and $\sigma$ on a case-by-case basis and are outside the scope of this paper.

Finally, it should be noted that the distributions discussed above are the distributions of the PDM and AOV statistics under a given trial period. In practice, many trial periods are tested and the period that minimizes the PDM or maximizes the AOV statistic is taken as the estimated period in the data. The FAP of the minimum PDM or maximum AOV statistic, however, does not follow these distributions and would need to be derived in a similar manner as the FAP for the LS power; this is left for future work. However, we reiterate here that the distributions of the minimum PDM and maximum AOV statistics, and their corresponding FAPs -- once derived -- will automatically control for ``testing" a range of trial periods, and multiple-testing is not a concern in that case.

\subsection{String-Length Method}

Another popular phase-based method is the SL method suggested by \cite{Burke1970} which measures the fit of the trial period $p$ based on the Euclidean distance between neighboring time-flux pairs after phase-folding. Specifically, the SL is defined as
\[
L := [(X_{t_1} - X_{t_T})^2 + (\phi_{t_1} - \phi_{t_T} + 1)^2]^{1/2} + \sum_{t=t_2}^{t_T} [(X_t - X_{t-1})^2 + (\phi_t - \phi_{t-1})^2]^{1/2}.
\]

Note, since the desired function is periodic, the distance between the first and last points is also included (as the first term) with an adjustment for the difference in phase values. The period $p$ that minimizes this distance is taken as the estimated period for the observed data values. \cite{renson1978method} criticized this method because the observed flux or magnitude values have different units from the phase values. To mitigate this issue under the assumption of Keplerian variation, \cite{Dworetsky1983} suggested scaling the observed values as $\Tilde{X}_t = (X_t - X_{\min})/2(X_{\max} - X_{\min}) - 0.25$. Under our $H_0$ in equation (\ref{eq:ht}), we can achieve homogeneity and ensure the SL statistic is robust to changes in units by scaling the observed values as $\Tilde{X}_t = (X_t - X_{\min})/(X_{\max} - X_{\min})$, which bounds both $\phi$ and $\Tilde{X}_t$ to be between zero and one.

When the observations are deterministic and from a continuously differentiable function, then $L$ can be determined using arc length equations from geometry. However, in the statistical setting, we must start with the distribution of the observed values. Specifically, under $H_0$, it is easy to show using standard transformation of random variable techniques that $Y_t = [(\Tilde{X}_t - \Tilde{X}_{t-1})^2 + (\phi_t - \phi_{t-1})^2]^{1/2}$ has a closed form distribution, see \cite{casella2021statistical} for these techniques. Unfortunately, however, the sum of these variables does not have a closed form distribution. Because of this difficulty, simulations or bootstrapping methods, with potential modifications as mentioned above, are needed to determine the FAP of the period that minimizes $L$.

\section{Non-Phase-Based Methods} \label{sec:npb_methods}

\subsection{Lomb-Scargle Power}\label{sec:ls}

The phase-based methods discussed in Section \ref{sec:pb_methods} are designed to work for time domain values of periodic functions, i.e. observations of the form (\ref{eq:periodx}). Alternatively, the \textit{LS power} examines the data for a periodic component in the frequency domain using the Fourier transform, i.e. observations of the form (\ref{eq:freq_domain}). The square of the Fourier transform is referred to as the {\em power spectral density} and is a measure of how much each frequency characterizes a periodic function \citep{Priestley1981, VanderPlas2018}. Knowing the power spectral density of a function is equivalent to knowing the periodic function since any periodic function can be reconstructed as a sum of sines and cosines with the frequencies from its power spectral density \citep{Priestley1981}. The LS power aims to estimate the power spectral density when the data are unevenly sampled in time (i.e. when classical time series methods do not apply).  Specifically, the LS power for trial frequency $f$ is 
\[ 
P(f) := \frac{1}{2} \bigg(\frac{\{\sum_{t = t_1}^{t_T} X_t \cos[2 \pi f (t - \tau)]\}^2}{\sum_{t = t_1}^{t_T} \cos^2[2 \pi f (t - \tau)]} + \frac{\{\sum_{t = t_1}^{t_T} X_t \sin[2 \pi f (t - \tau)]\}^2}{\sum_{t = t_1}^{t_T} \sin^2[2 \pi f (t - \tau)]} \bigg),
\]
where $\tau := \frac{1}{4 \pi f} \tan^{-1}\bigg[\frac{\sum_{t = t_1}^{t_T} \sin (4 \pi f t)}{\sum_{t = t_1}^{t_T} \cos (4 \pi f t)} \bigg]$ \citep{Lomb1976, Scargle1982, VanderPlas2018}.  When the data are generated from equation (\ref{eq:periodx}), $2P(f)/\sigma^{2} \sim \chi_2^2$ for each $f$ which is established by Cochran's theorem and the fact that $\sum_t \sin[2\pi f(t- \tau)] \cos[2\pi f(t - \tau)] = 0$, for any $f$. Further, it can be shown that $P(f)/\sigma^{2} \sim \exp(1)$ where $\exp(1)$ is the exponential distribution with scale parameter one \citep{casella2021statistical, Scargle1982, VanderPlas2018}. Finally, because having data with mean zero and known $\sigma^2$ is rare, the LS power is often calculated with data that has been centered around the empirical mean and scaled using the empirical variance. We refer to $2P(f)/\sigma^{2}$ using these empirical values as the scaled LS power.

One of the main difficulties of period detection using the LS power is determining the significance of the largest peak. This difficulty is because the power at one frequency is correlated with the power at other frequencies through a convolution of the true signal and the observation times, where a convolution is defined as $f * g(t) := \int_{-\infty}^{\infty} f(\tau)g(t-\tau)d\tau$ \citep{VanderPlas2018}. In the case of Gaussian errors as in equation (\ref{eq:periodx}), \cite{Baluev2008} adapted theory (based on work from \cite{Davies1977}, \cite{Davies1987}, and \cite{Davies2002}) for approximating the distribution of the maximum of chi-squared random variables over frequencies $f$, for the largest LS power. While this approximation gives an analytic result that can be easily implemented, it is an upper bound and can greatly overestimate the type 1 error rate when aliasing occurs \citep{VanderPlas2018}. 

\begin{wrapfigure}{l}{0.55\textwidth}
    \begin{minipage}{0.5\textwidth}
    \includegraphics[width = \textwidth]{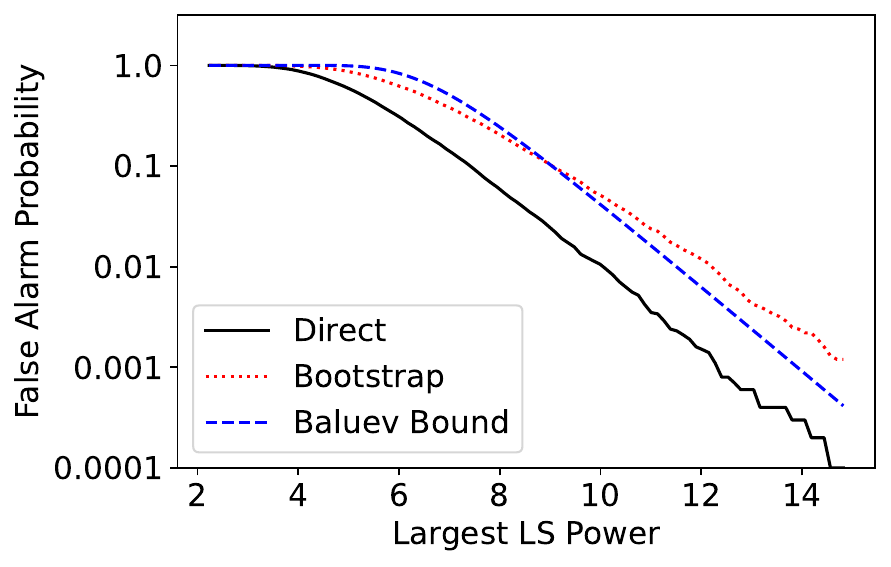}
    \captionof{figure}{Comparison of the false alarm probabilities calculated directly and using bootstrap resampling versus the theoretical bound provided by \cite{Baluev2008} for the largest observed LS power.}\label{fig:fap_bound}
    \end{minipage}
\end{wrapfigure}

In Figure \ref{fig:fap_bound}, we examine the type 1 error rate of the maximum LS power compared to the upper bound derived in \cite{Baluev2008}. We follow \cite{VanderPlas2018} and generate 100 unevenly sampled observations $m_{1}, \dots, m_{100}$ from the set $\{0, 1, \dots, 5 \cdot 24 \}$ (i.e. 100 observations over 5 days) and rescale the observation times as $t_i = 0.01\cdot m_i$ for $i \in \{1,\dots,100\}$. Following \cite{VanderPlas2018}, we generate a grid of frequency values from $1/(24 \cdot 25)$ to $24 \cdot 10$. For 10,000 iterations, we randomly generate the flux values from a Gaussian distribution with $\sigma^2 = 1$ (i.e. under the null hypothesis) and calculate the LS power at each point in the frequency grid. These datasets are used to calculate the FAP directly. Additionally, we create a bootstrapped version of one of the datasets and calculate the corresponding LS power for the same frequency grid for 10,000 iterations. These bootstrapped datasets are used to calculate the FAP using bootstrapping methods that would be necessary in practice without an analytical bound. Specifically, the direct and bootstrap FAPs are calculated as follows. First, we create a grid over the range of observed maximum LS power values. Then for each grid point, we calculate the percentage of the observed maximum LS power values that are larger than it. Finally, the Baluev bound is calculated using equation (6) of \cite{Baluev2008}.

\section{Simulation Study} \label{sec:sims}

The purpose of our simulation study is to demonstrate the properties of the classical period detection statistics discussed in this paper predominately under $H_0$. How these properties change under heteroscedastic error and a particular case in which $H_a$ is true are explored as well but are for demonstration purposes. Exhaustive power analyses can be conducted for detecting a non-constant periodic function in the data using these statistics but would need to be performed based on the signal and noise levels expected for a given application. Thus, further collaborations and research is needed to conduct such power analyses. Each subsection below details how the data were generated and examines the properties of the statistics discussed above under $H_0$ with homoscedastic error, $H_0$ with heteroscedastic error, and $H_a$ with homoscedastic error. For all scenarios, we followed \cite{mowlavi2022gaia} and used a frequency range between 0.005 and 15 $d^{-1}$ (a period range of 1.6 hours to 200 days) with a fixed frequency step of $10^{-5} d^{-1}$. We chose this grid because our simulation study under $H_a$ is based on one of the models fit to a binary star system found in Gaia data by the same group.

\subsection{Properties of Statistics and Periodograms under \texorpdfstring{$H_0$}{TEXT}}\label{sec:simh0}

To demonstrate the properties of the PDM, AOV, SL, and LS power statistics under $H_0$ with homoscedastic error, we generate 300 simulated datasets based on observed eclipsing binary data from the Gaia telescope. Our generation of simulated data can be decomposed into three main steps. First, we identify Gaia sources that have been labeled as eclipsing binaries by \cite{mowlavi2022gaia}. Second, we determine the average observed flux and average observed flux error for these eclipsing binaries and generate simulated $G$-magnitudes under $H_0$ and corresponding errors based on these observed values. Finally, the observation times of the simulated data are generated based on actual observation times from a randomly selected source from Step 1.

We perform Step 1 as follows. First, we keep Gaia observations with $G$-magnitudes between 18.3 and 19.5; these apparent magnitudes were chosen because they approximately cover the range of $G$-magnitudes that are most likely to be observed and are the range of magnitudes that will be simulated under $H_1$ \citep{mowlavi2022gaia}. We then merge this data with the data used in \cite{mowlavi2022gaia} to obtain sources that have been labeled as eclipsing binaries.

Step 2 is needed because the original observed weighted-mean flux values from Gaia have heteroscedastic error and the data under $H_0$ is assumed to have homoscedastic error. We calculate a reasonable homoscedastic error by determining the average flux error of the observations from Step 1 where we define the average flux error as the product of the average observed weighted-mean flux and the average flux error ratio (observed flux error divided by observed flux). We denote this new average flux error as $\sigma_f$. The simulated weighted-mean flux values, $f$, are then generated as random draws from a normal distribution with the mean equal to the average observed weighted-mean flux and standard deviation $\sigma_f$. The simulated $G$-magnitudes, $G$, are then calculated as
\begin{equation}\label{eq:gfromf}
    G = - 2.5\log_{10}(f) + G_0
\end{equation} where $G_0 = 25.6874$ \citep{carrasco2016gaia, riello2021gaia}. Additionally, using the Delta Method for the first term in equation (\ref{eq:gfromf}) \citep{casella2021statistical}, the standard deviation of $G$ is approximately
\begin{equation}
    \sigma_G = \sqrt{\bigg(\frac{2.5\sigma_f}{f \ln 10}\bigg)^2 + \sigma_{G_0}^2}
\end{equation} where $\sigma_{G_0} = 0.0028$ is the zero point uncertainty from \cite{riello2021gaia}. The final simulated $G$-magnitudes are then set as $G/\sigma_G$, which we refer to as the scaled $G$-magnitudes. In other words, the scaled $G$-magnitudes generated in this way correspond to the $X_t$ in equation (\ref{eq:periodx}) with a constant periodic signal, i.e. under $H_0$, and (approximately) homoscedastic noise. We generate 43 such simulated $G$-magnitudes because that was the average number of observations a source had in our merged data from Step 1.

Finally, the observation times for the simulated data under $H_0$ are set. To do this, we randomly selected a source from Step 1 with 43 observations. For the evenly-spaced observation scenario, we generated a grid of observation times between the first and last time stamps of the selected source. Otherwise, for the unevenly-spaced observation scenario, we set the observation times as the actual observation times from the source. Figure 3 shows the scaled $G$-magnitudes with their corresponding observation times used for calculating the classical period-detection statistics discussed in Sections \ref{sec:pb_methods} and \ref{sec:npb_methods}.

\begin{figure}
    \centering
    \includegraphics[scale = 0.5]{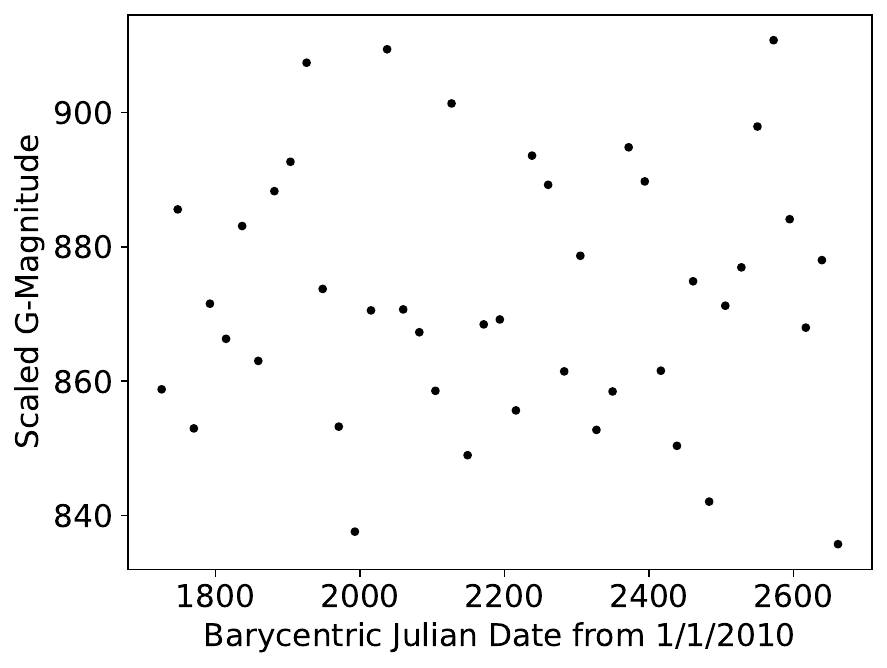}
    \caption{Simulated observed scaled $G$-magnitude under the null hypothesis that observations are Gaussian noise. Only the time stamp differs in the even- and unevenly-spaced observation cases. Note, the scale of the $G$-magnitude is different from typically observed values due to dividing by $\sigma_G$.} 
    \label{fig:obs_null}
\end{figure}

\begin{figure}
    \centering
    \includegraphics[scale = 0.27]{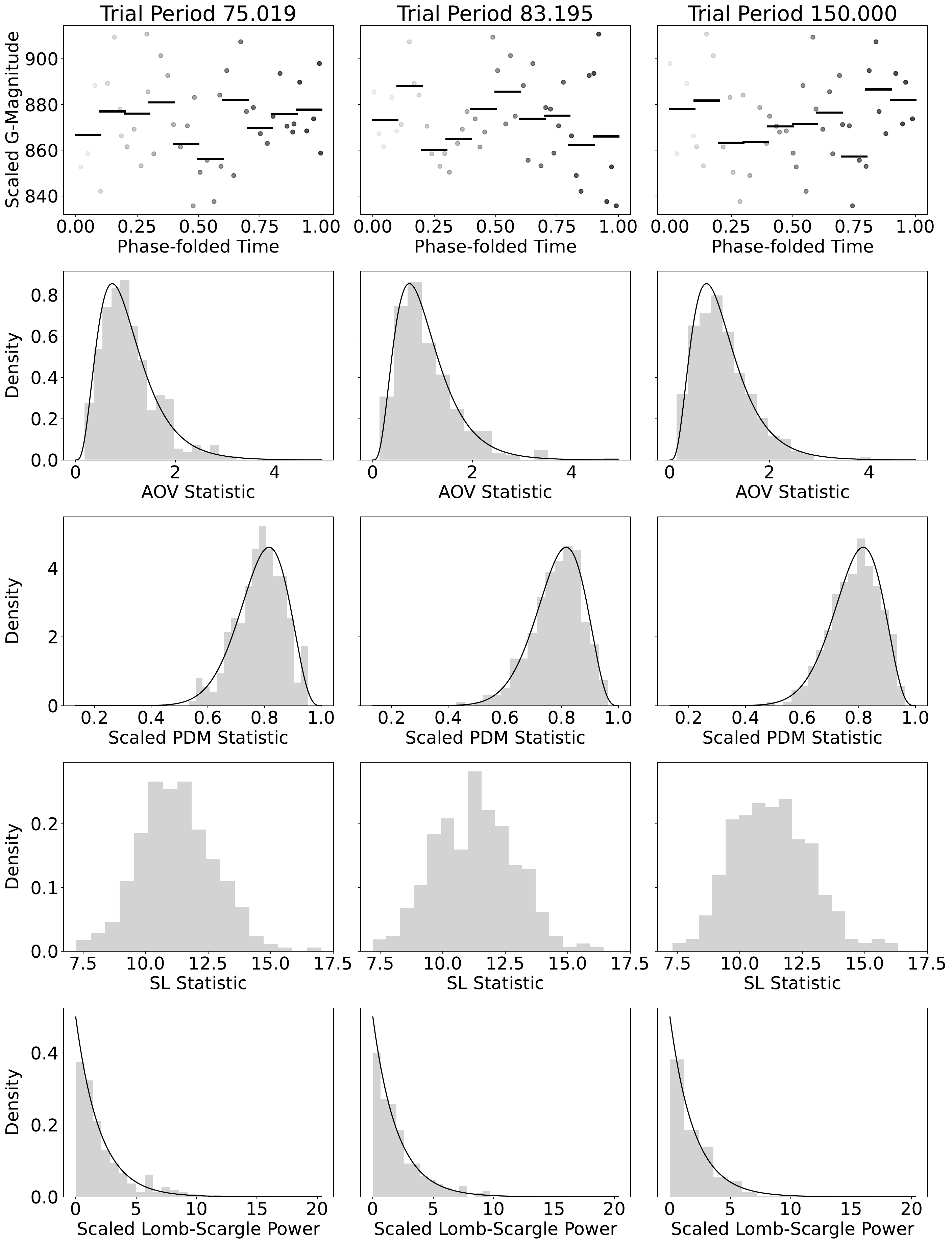}
    \caption{The first row displays the simulated phase-folded, scaled $G$-magnitude values generated as independent and identical Gaussian realizations evenly spaced in time. The observed bin means are indicated by the black lines and the observations within each bin are the same shade of grey. The second, third, fourth, and fifth rows display histograms of the AOV, scaled PDM, and SL statistics and scaled LS power, respectively. The solid black line is the density of the central F distribution in the second row, the central beta distribution in the third row, and the chi-squared distribution with two degrees of freedom in the fifth row. As the SL statistic does not have a closed-form density function, none is plotted here. The plotted periods are chosen for comparison to the case where the data contain a non-constant periodic function in Section \ref{sec:h1}.}
    \label{fig:null_even}
\end{figure}

\begin{figure}
    \centering
    \includegraphics[scale = 0.27]{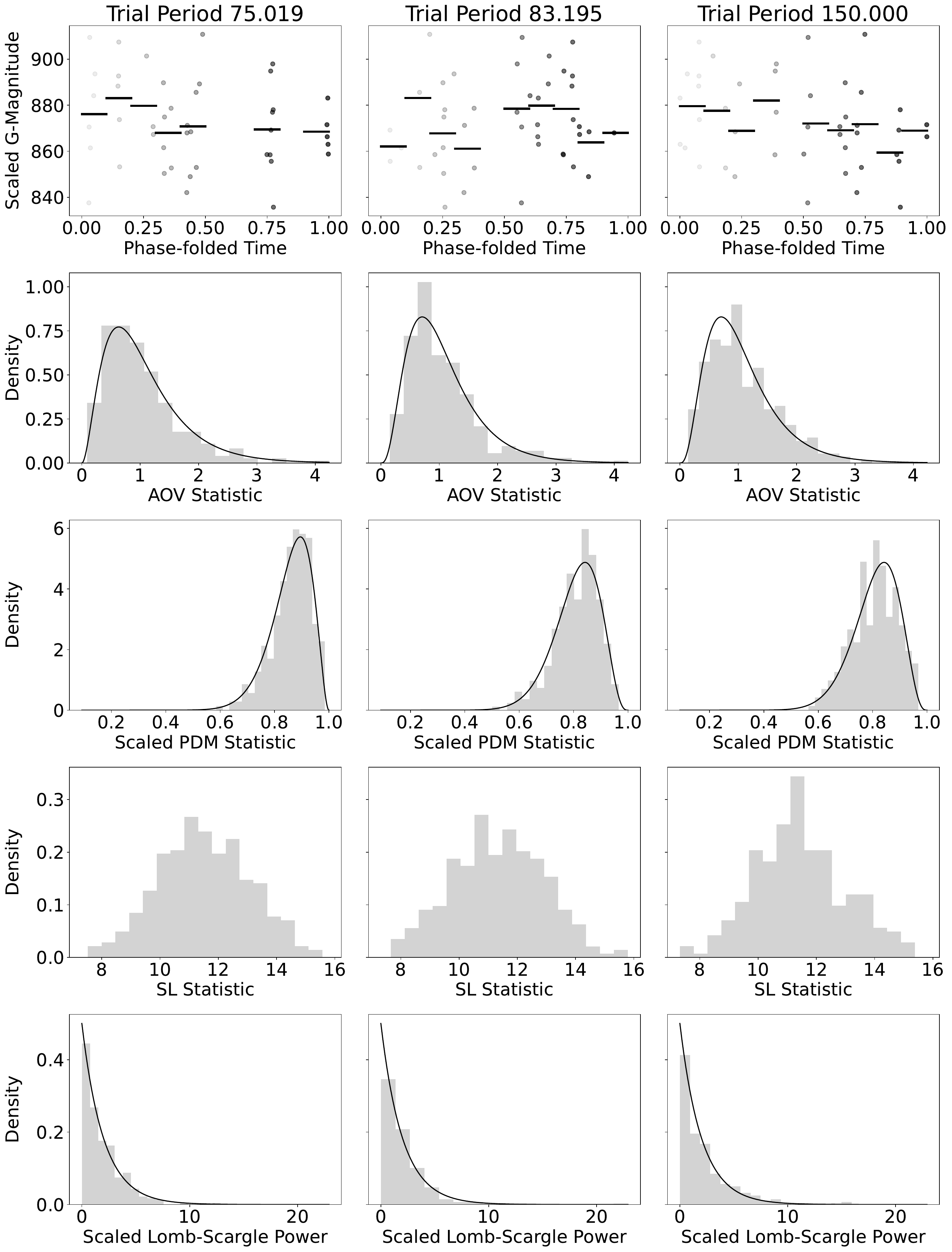}
    \caption{Same as Figure \ref{fig:null_even} except the observed data are unevenly spaced in time.}
    \label{fig:null_uneven}
\end{figure}

Figures \ref{fig:null_even} and \ref{fig:null_uneven} show the phase-folded data with histograms of the corresponding AOV, scaled PDM, SL, and scaled LS power statistics for evenly- and unevenly-spaced observation times, respectively. The solid black line is the density of the central F distribution for the AOV statistics, the density of the central beta distribution for the scaled PDM statistics, and the density of the chi-squared distribution with two degrees of freedom for the scaled LS power. The plotted periods are chosen for comparison to the case where the data contain a non-constant periodic function in Section \ref{sec:h1}. As demonstrated by Figures \ref{fig:null_even} and \ref{fig:null_uneven}, the AOV, PDM, and LS power statistics follow their respective theoretical distributions under $H_0$ as shown in Sections \ref{sec:pb_methods} and \ref{sec:npb_methods}. Not displayed, however, is that the scaled PDM statistic calculation breaks if all of the phase-folded times are in the same bin. This can and does occur when the observation times are evenly-spaced but does not occur in the unevenly-spaced case. When this happened, we set the scaled PDM to be missing.

\begin{figure}
    \centering
    \includegraphics[scale = 0.19]{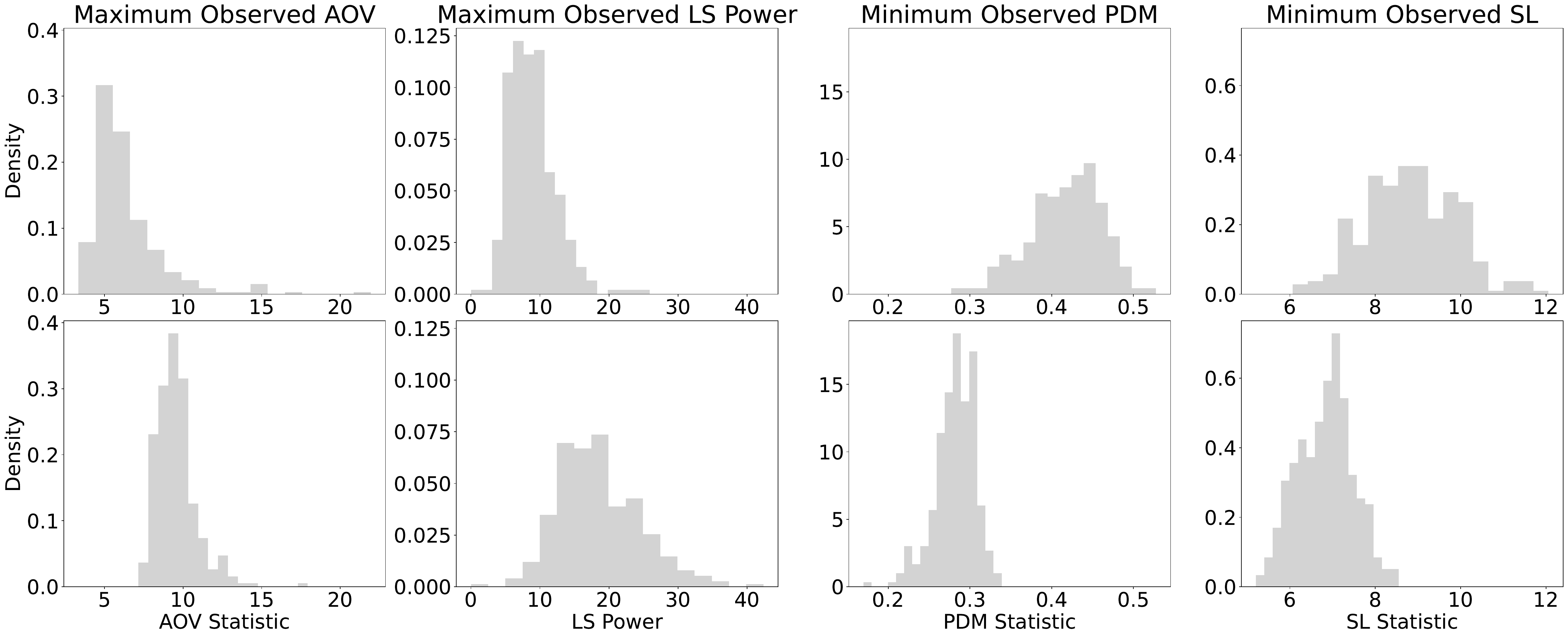}
    \caption{Maximum AOV statistic and scaled LS Power and minimum scaled PDM and SL statistics for data generated under $H_0$ with evenly- (top) and unevenly- (bottom) spaced observations. Note, we have excluded the densities under the null displayed in Figures \ref{fig:null_even} and \ref{fig:null_uneven} because the empirical densities here are very different from those for a single trial period/frequency. Specifically, note the changes in the axes values.}
    \label{fig:xvals_pas}
\end{figure}

Figure \ref{fig:xvals_pas} shows the maximum AOV and scaled LS power and minimum scaled PDM and SL statistics. Clearly, the distributions of the extreme values do not follow the theoretical distributions under $H_0$ discussed in Sections \ref{sec:pb_methods} and \ref{sec:npb_methods}. As such, the FAP using these statistics would need to be calculated using their respective extreme value distributions, of which only the LS power has a known bound.

\begin{figure}
    \centering
    \includegraphics[scale = 0.21]{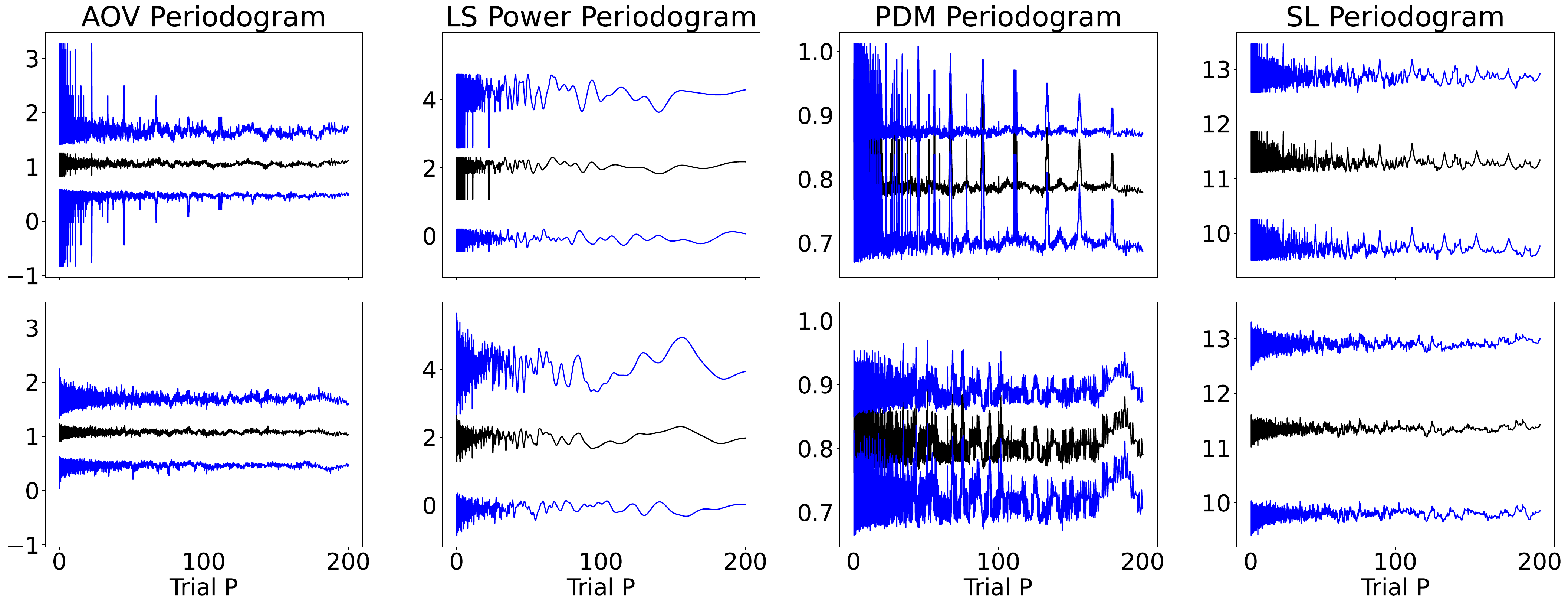}
    \caption{Periodograms of the AOV (left), LS power (middle-left), scaled PDM (middle right), and SL (right) statistics over a range of trial periods for evenly- (top) and unevenly- (bottom) spaced observation times under $H_0$ with homoscedastic error. The black line is the mean periodogram for 300 simulated data sets, at each trial period, and the blue lines are the mean periodogram plus and minus the Monte Carlo standard deviation, at each trial period. Note, the y-axes are the values of the respective statistics in the titles.}
    \label{fig:null_periodogram}
\end{figure}

Finally, Figure \ref{fig:null_periodogram} shows the periodograms of the AOV, scaled PDM, SL, and scaled LS power statistics over the range of trial periods. The black line is the mean periodogram at each trial period, and the blue lines are the mean periodogram plus and minus the Monte Carlo standard deviation at each trial period. The periodograms of the AOV, SL, and LS power statistics suggest the data come from a constant function since there is no clearly dominating peak. On the other hand, the periodogram of the scaled PDM has multiple peaks when the data have evenly-spaced observation times that suggest the data could come from a non-constant periodic function. This issue, however, is alleviated when the observation times are unevenly-spaced but is still concerning as additional resources would need to be expended to determine that the result is spurious.

\subsection{Robustness of Statistics and Periodograms under Heteroscedasticity}

\begin{figure}
    \centering
    \includegraphics[scale = 0.48]{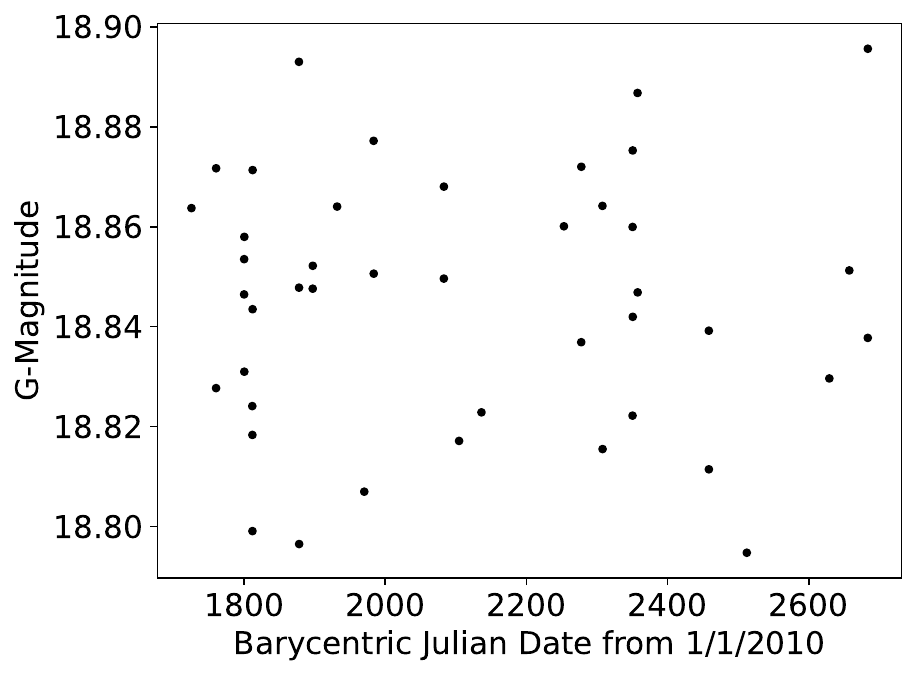} \hfill
    \includegraphics[scale = 0.48]{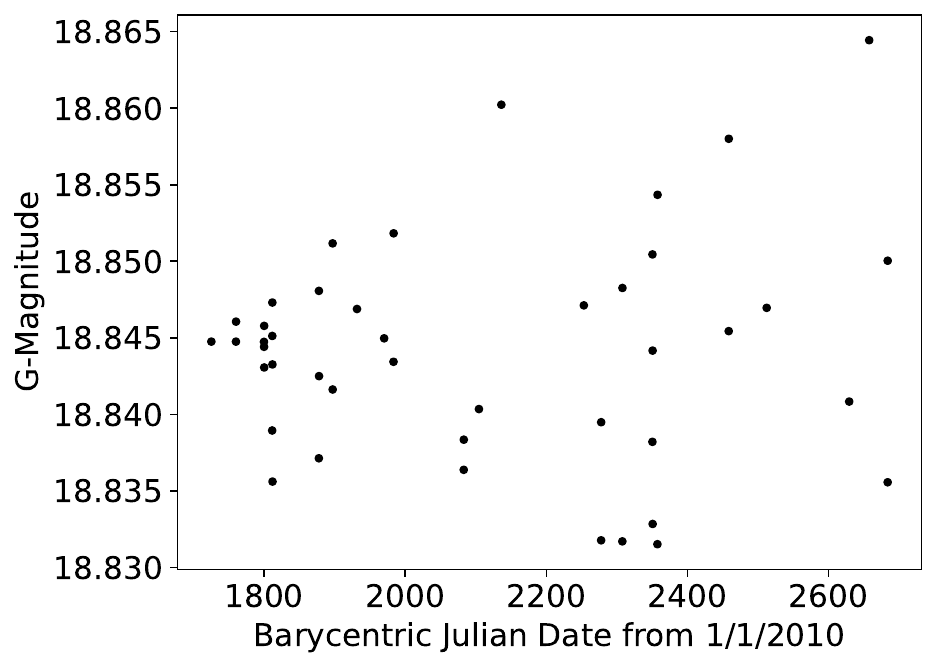}
    \caption{Observed data under null hypothesis with heteroscedastic error from not rescaling the observed $G$-magnitudes (left) and from increasing standard deviations as time increases (right).}
    \label{fig:obs_hetero}
\end{figure}

Now, we explore how robust the PDM, AOV, SL, and LS power values are under $H_0$ with heteroscedastic noise. We do this under two scenarios. In the first scenario, we generate the data exactly like how the data was generated in Section \ref{sec:simh0} but we do not rescale $G$ by $\sigma_G$. In the second scenario, we set the weighted-mean flux error to increase from the minimum observed flux error ratio to the maximum observed flux error ratio of the data in Step 1. Again, we do not rescale $G$ by $\sigma_G$. The second case aims to represent the degradation or mis-calibration of an instrument over time with the more reaslistic use case of the (unscaled) $G$-magnitudes, while the first case isolates the effect of not scaling the $G$-magnitudes. For both cases, the observation times are unevenly-spaced as described above. Figure \ref{fig:obs_hetero} shows the observed data under these two cases.

\begin{figure}
    \centering
    \includegraphics[scale = 0.27]{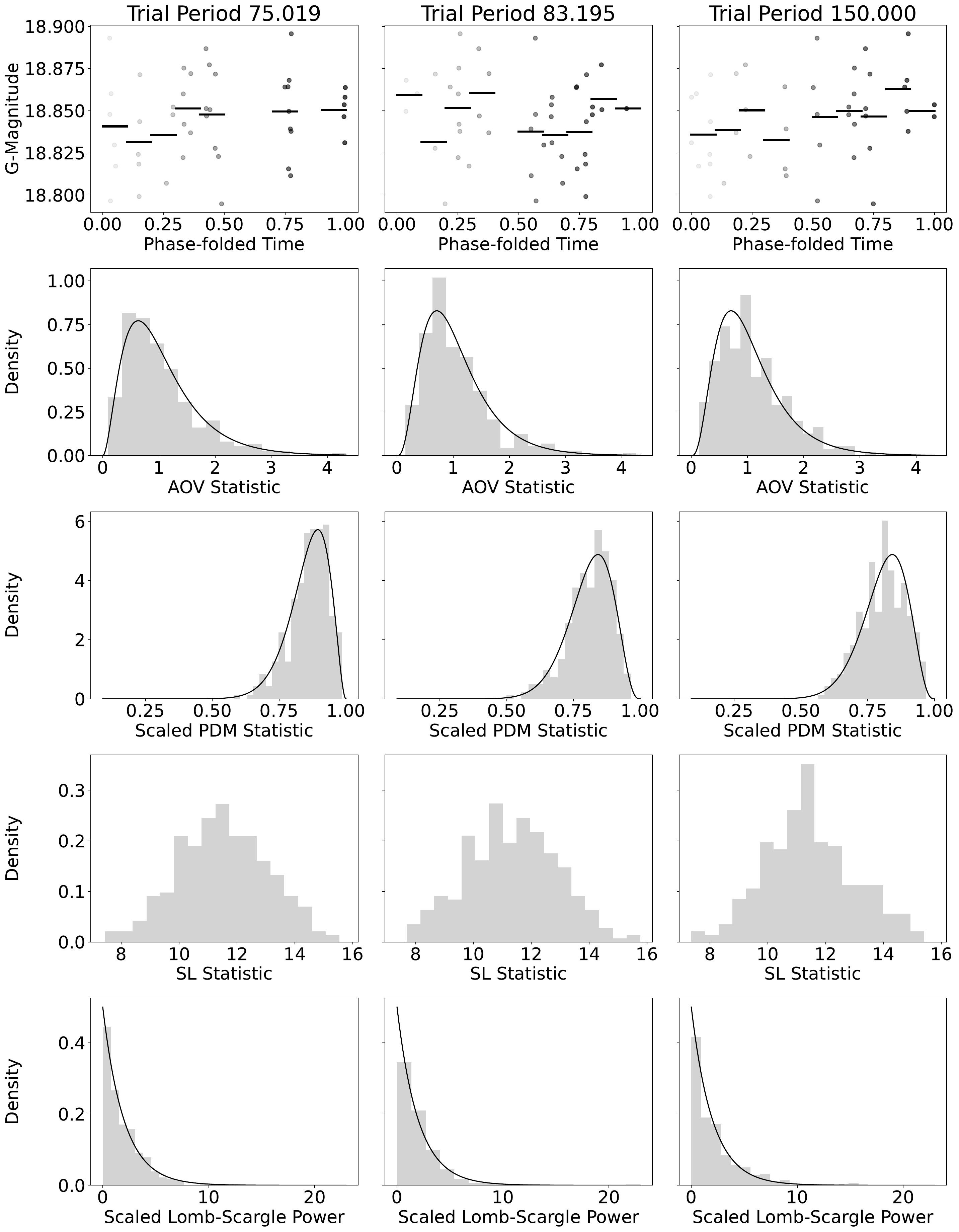}
    \caption{Same as Figure \ref{fig:null_uneven} but with generated data containing heteroscedastic noise from not rescaling $G$ by $\sigma_G$.}
    \label{fig:null_hetero_trans}
\end{figure}

\begin{figure}
    \centering
    \includegraphics[scale = 0.27]{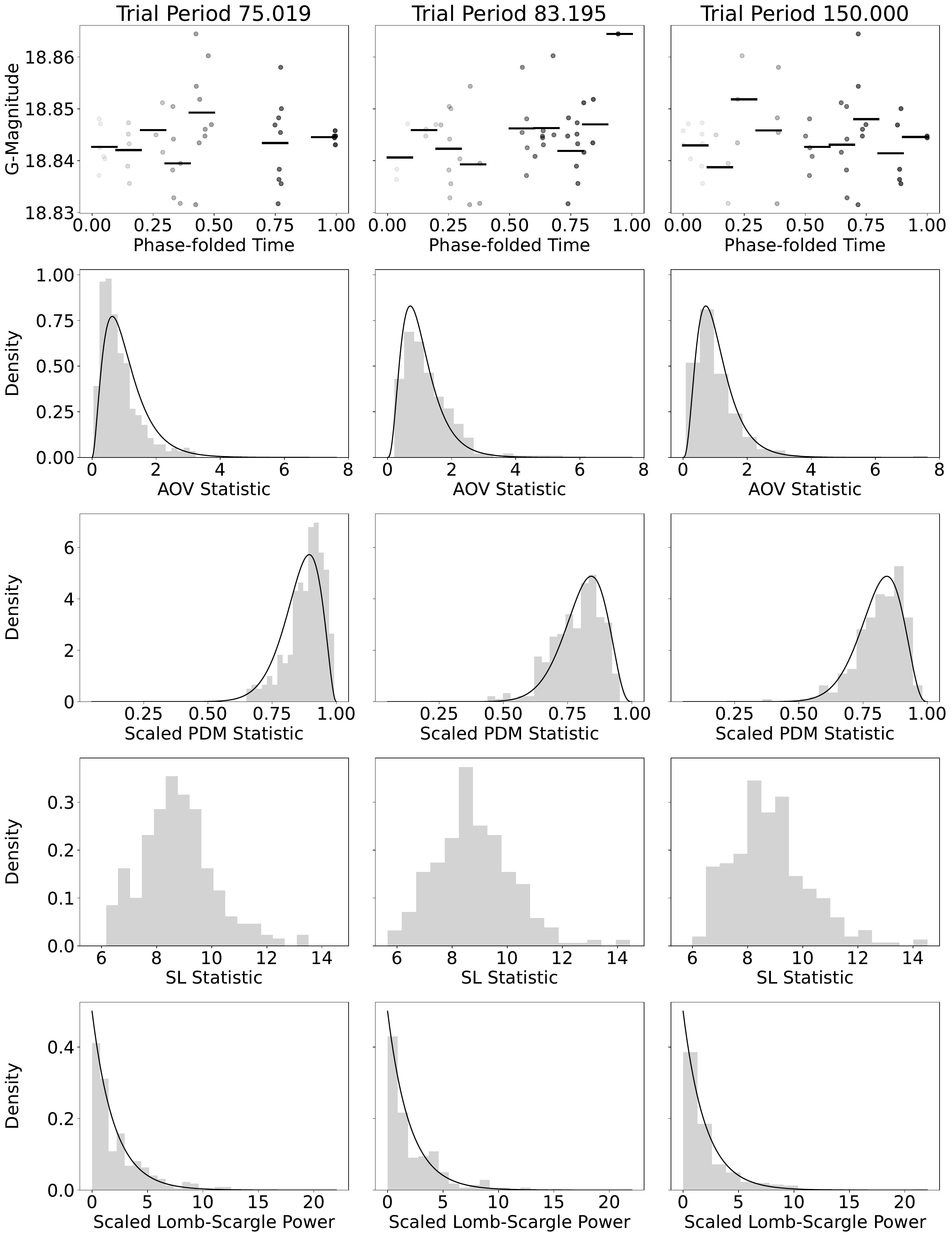}
    \caption{Same as Figure \ref{fig:null_uneven} but with generated data containing noise that increases over time.}
    \label{fig:null_hetero_sim}
\end{figure}

Figures \ref{fig:null_hetero_trans} and \ref{fig:null_hetero_sim} are the same as Figure \ref{fig:null_uneven} but for the heteroscedastic data generated for this section. Unsurprisingly, the AOV, scaled PDM, and scaled LS power statistics are clearly robust to the heteroscedasticity from not scaling $G$ by $\sigma_G$ in that they still follow their theoretical distributions under $H_0$; this is because $\sigma_G$ is also random (from the randomness in $f$) and does not depend on time. Similarly, the SL statistic appears to be robust to the heteroscedasticity introduced from not re-scaling $G$ as the histogram in Figure \ref{fig:null_hetero_trans} closely resembles that from Figure \ref{fig:null_uneven}. Surprisingly, however, the AOV, scaled PDM, and scaled LS power statistics are fairly robust to increasing noise over time. This is likely due to the phase-folding shuffling the observations so that the observations appear to have homoscedastic noise. Although, for the trial period of 75.019 days the AOV statistic appears to decrease while the PDM statistic appears to increase; this is not terribly concerning, though, as these values are unlikely to result in a false alarm. However, the SL statistic does not appear to be robust to increasing noise over time as it appears to decrease noticeably from the values observed in Figure \ref{fig:null_uneven}. This shift is concerning as smaller values of the SL statistic may result in a false alarm.

\begin{figure}
    \centering
    \includegraphics[scale = 0.21]{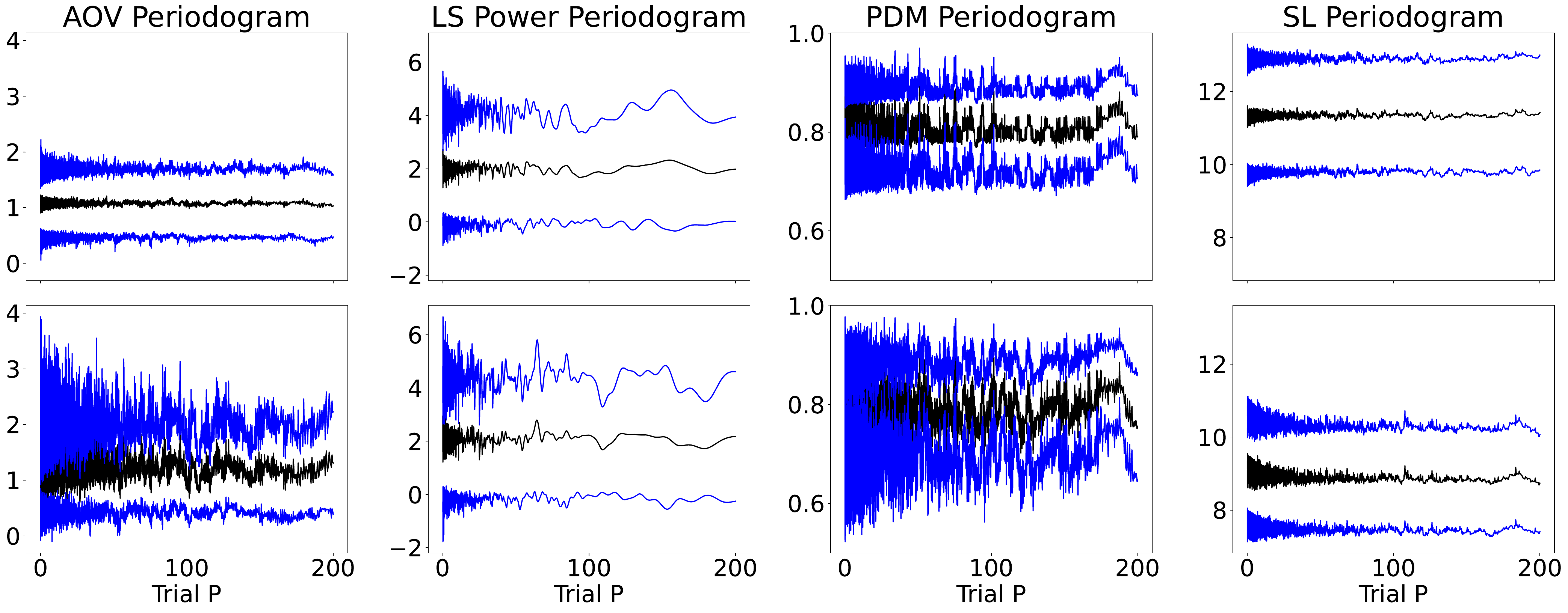}
    \caption{Same as Figure \ref{fig:null_periodogram} except top row is from data with heteroscedastic error from not rescaling $G$ by $\sigma_G$ and bottom row from data containing noise that increases over time.}
    \label{fig:hetero_periodogram}
\end{figure}

Finally, Figure \ref{fig:hetero_periodogram} is the same as Figure \ref{fig:null_periodogram} except for the data generated in this subsection. Interestingly, the spurious results in the scaled PDM periodograms discussed in the previous subsection disappear with the addition of heteroscedasticity. All of the periodograms suggest the data come from a constant function.

\subsection{Power Analysis of Statistics and Periodograms under \texorpdfstring{$H_a$}{TEXT}}\label{sec:h1}

Finally, we explore how the PDM, AOV, SL, and LS power statistics perform under a particular example in which $H_a$ is true, i.e. a case in which the data come from a non-constant periodic function. To generate this data, we chose one of the fitted models from \cite{mowlavi2022gaia} that had well separated fitted Gaussians which mimicked a well-detached binary system. We chose this type of model as this would be the easiest case to determine the periodic component of the data and thus demonstrate the ability, the power in statistical terms, of these statistics to detect a non-constant period in a realistic data scenario.

\begin{figure}
    \centering
    \includegraphics[scale = 0.5]{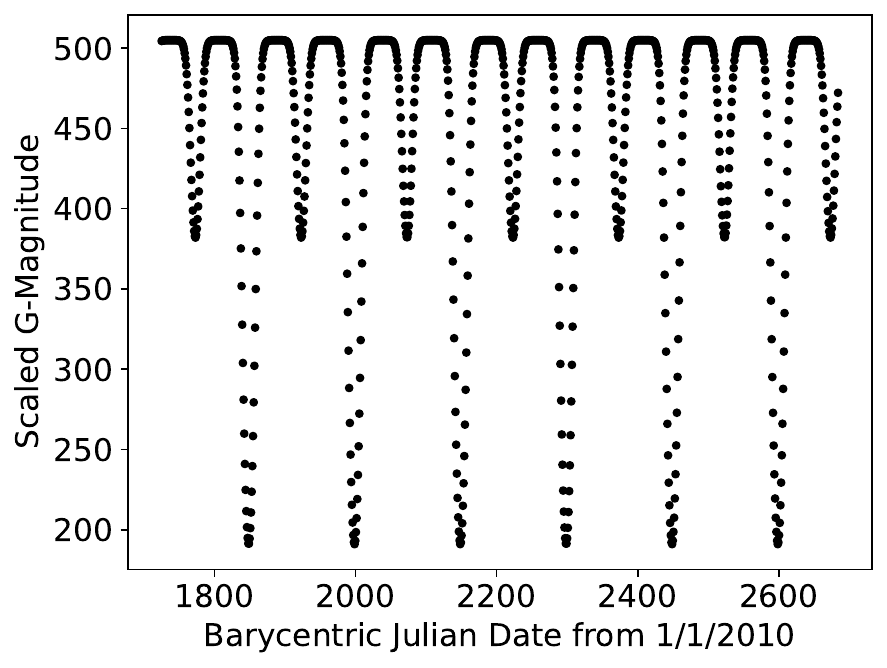} \hfill
    \includegraphics[scale = 0.5]{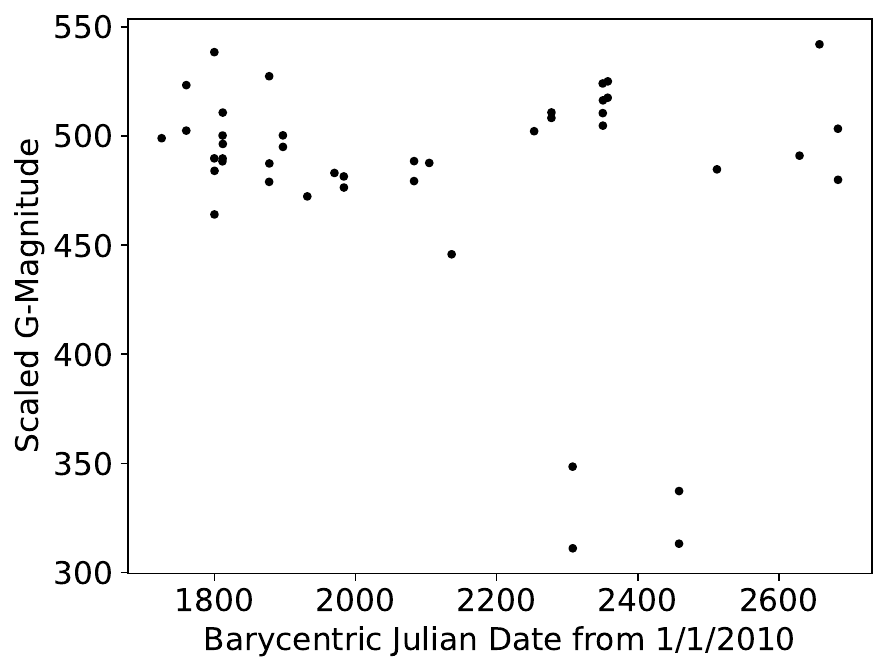}
    \caption{Underlying periodic function (left) and observed data (right) under $H_a$. Note, the scale of the $G$-magnitude is different from typically observed values due to dividing by $\sigma_G$.}
    \label{fig:obs_h1}
\end{figure}

We arbitrarily set a period of 150 days for the binary system and set the weighted-mean flux as the inverse of equation (\ref{eq:gfromf}) where $G$ is determined from the \cite{mowlavi2022gaia} model. We then generated 300 datasets by randomly drawing $f$ from a normal distribution with the new weighted-mean flux as the mean and $\sigma_f$ from Section \ref{sec:simh0} as the standard deviation, converting $f$ to $G$ using equation (\ref{eq:gfromf}), and rescaling the simulated $G$-magnitudes to be $G/\sigma_G$. Finally, we set the observation times to be unevenly-spaced as described above. Figure \ref{fig:obs_h1} shows the underlying periodic function and the simulated values analyzed in this subsection.

\begin{figure}
    \centering
    \includegraphics[scale = 0.27]{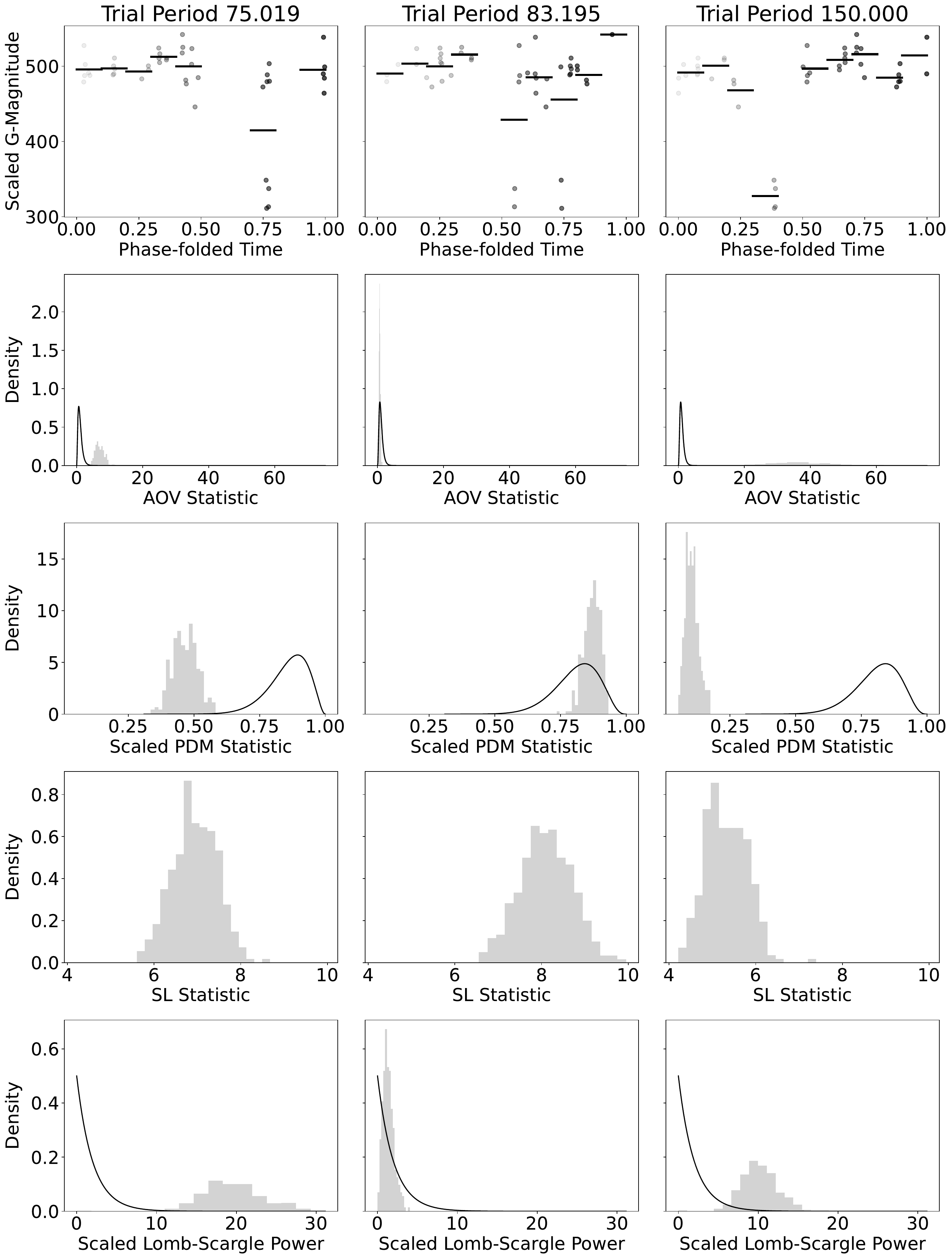}
    \caption{Same as Figure \ref{fig:null_uneven} but with generated data from a non-constant periodic function with homoscedastic noise and $p_0 = 150$.}
    \label{fig:h1}
\end{figure}

\begin{figure}
    \centering
    \includegraphics[scale = 0.185]{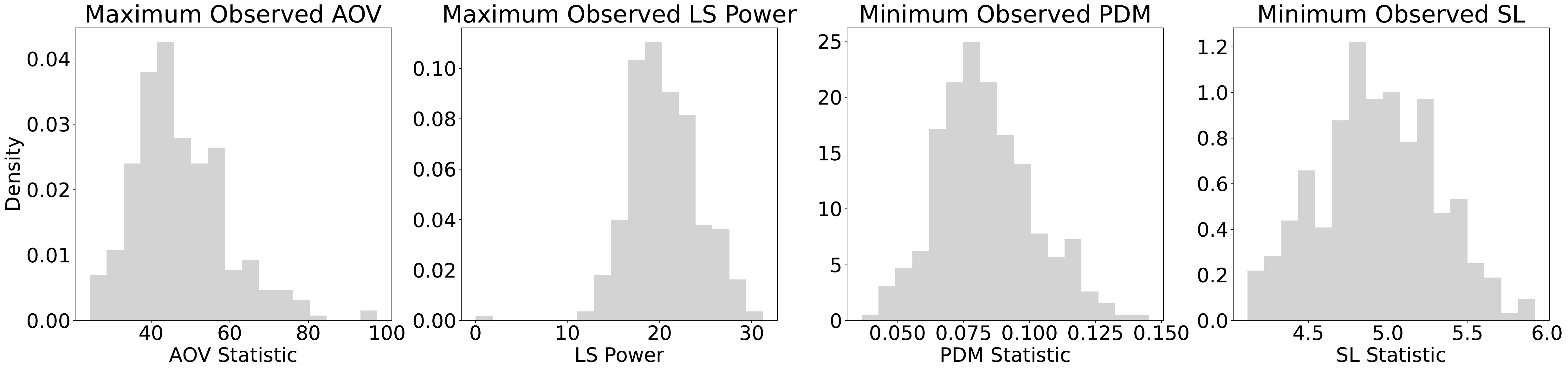}
    \caption{Maximum AOV statistic and LS Power and minimum scaled PDM and SL statistics for data generated under $H_a$ with evenly- (top) and unevenly- (bottom) spaced observations. Note, we have again excluded the densities under the null.}
    \label{fig:xvals_h1}
\end{figure}

Figure \ref{fig:h1} is the same as Figure \ref{fig:null_uneven} but for the data generated as under $H_a$ for this section. Clearly, none of the statistics follow their distributions under $H_0$ as all of the observed values are in the tails of these distributions. Finally, Figure \ref{fig:xvals_h1} shows the maximum AOV and scaled LS power and minimum scaled PDM and SL statistics for this subsection's data. Comparing with Figure \ref{fig:xvals_pas}, it is clear the maximum observed AOV and minimum scaled PDM and SL statistics using data from a non-constant periodic function differ from those corresponding statistics using data generated from a constant function. Oddly, this does not appear to be the case for the LS power. Further, the AOV statistic seems to be the most sensitive to detecting a periodic function because the observed maximum values are fairly far away from those observed under $H_0$. In fact, the difference between the smallest observed maximum under $H_a$ and largest observed maximum under $H_0$ is about 6. The PDM statistic seems the next most sensitive to detecting a (non-constant) periodic signal because the observed minimums under $H_a$ are well-separated from those observed under $H_0$; however, it is possible that some of the values observed under $H_0$ may be in the tail of the distribution under $H_a$ but are unobserved due to only performing 300 simulations. In fact, the difference between the smallest observed minimum under $H_0$ and the largest observed minimum under $H_a$ is about 0.024. Finally, based on the simulations the SL statistic seems the least sensitive to detecting a (non-constant) periodic signal for the same reason as the PDM statistic and that the observed minimums under $H_a$ actually appear as observed minimums under $H_0$. These simulations agree with the observation made in \cite{Feigelson2021} that while the SL method avoids the arbitrary binning of observations, it is less sensitive to detection of the true period in practice.

\begin{figure}
    \centering
    \includegraphics[scale = 0.22]{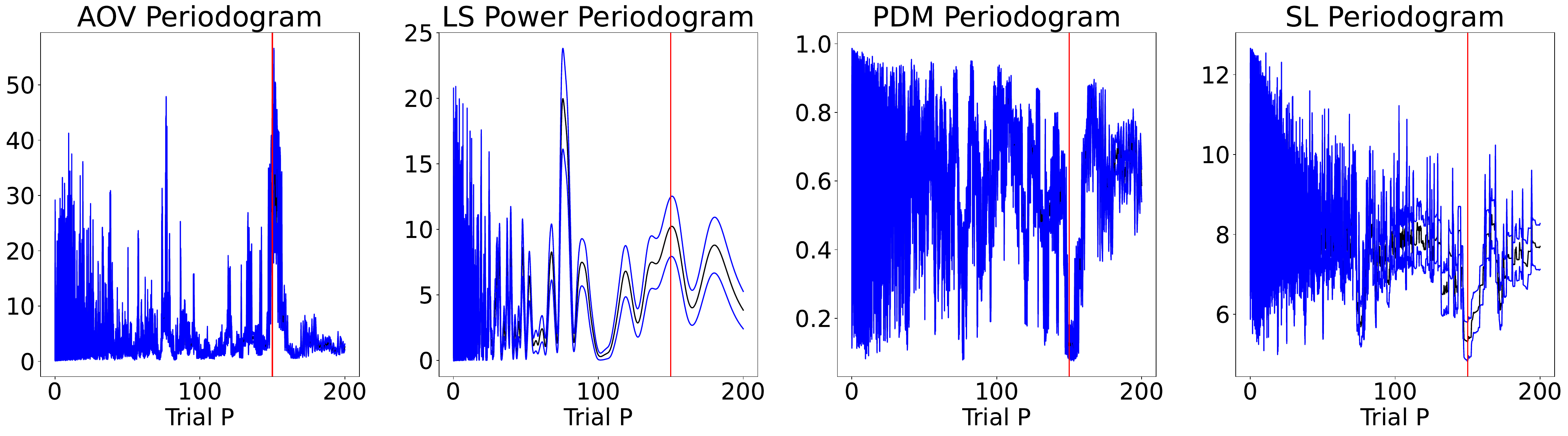}
    \caption{Same as Figure \ref{fig:null_periodogram} except data is generated from a non-constant periodic function with homoscedastic noise, i.e. under a potential $H_a$, with true period $p_0 = 150$ (vertical line).}
    \label{fig:h1_periodogram}
\end{figure}

Finally, Figure \ref{fig:h1_periodogram} shows the periodograms of the AOV, PDM, and SL statistics and the LS power values for the data generated for this subsection. All of the periodograms have a strong peak at a false period of 75 days, which is due to both of the dips in Figure \ref{fig:obs_h1} aligning as can be seen in the top-left panel of Figure \ref{fig:h1}. This is a common issue for periodograms \citep{hara2023statistical}. Interestingly, all of the phase-based methods have a stronger peak at the correct period of 150 days but the LS power periodogram does not. Finally, the AOV statistic appears to have smaller fluctuations at incorrect periods compared to the PDM and SL statistics.

\section{Concluding Remarks} \label{sec:conc}
Detecting a periodic signal in data is a historic and on-going area of research for astronomers. Much research has been conducted on and using the PDM \citep{Stellingwerf1978}, AOV \citep{Schwarzenberg1989}, and SL \citep{Dworetsky1983} statistics, and the LS power values \citep{Lomb1976, Scargle1982} by astronomers. However, there has been little input from statisticians which has created concerns about the theoretical and empirical properties of these methods \citep{Feigelson2021}. Our paper aims to fill this gap. We confirm that a scaled version of the PDM statistic follows a beta distribution, the AOV statistic follows an F distribution, and the scaled LS power follows a chi-squared distribution with two degrees of freedom when the data are perturbations of a constant function. However, the SL statistic does not have a closed-form distribution. We emphasize that calculation of the FAP, or p-value in the statistics literature, can only use these distributions if one is only concerned about the FAP for a particular trial period. However, as demonstrated by Figure \ref{fig:xvals_pas} the FAP would need to be calculated under a different distribution if an extreme value such as the minimum or maximum is used to test for a non-constant periodic signal. In the case of the SL statistic, a modification of the bootstrap method would need to be employed for any calculation of the FAP, e.g. those suggested by \cite{fukuchi1994bootstrapping}. Finally, as demonstrated by Figure \ref{fig:xvals_h1}, we find the AOV statistic is most sensitive to a non-constant periodic signal while the SL statistic and LS power are the least sensitive. Additionally, the phase-based methods are less sensitive to false periods, particularly the AOV statistic, compared to the LS power. These results align with the conclusions drawn by \cite{Heck1985}.

However, our simulation study focuses on the properties of these statistics under the null hypothesis that the data are fluctuations from a constant function and performed little exploration on the power of these statistics to detect a periodic signal in data. Further collaboration and research is needed to perform power analyses (how well these statistics detect periodic signals) that astronomers would find useful, such as those discussed in \cite{Heck1985}. Finally, additional research is needed to identify ways to overcome challenges such as faint signals, irregular observing cadences, and interactions between observations and noise \citep{Feigelson2021, hara2023statistical}.

\section*{Acknowledgements}

This work was partially supported by National Science Foundation grant NSF-AST 2206853. Additionally, the authors would like to acknowledge the computing resources provided by North Carolina State University High Performance Computing Services Core Facility.

\newcommand{\newblock}{}
\bibliographystyle{agsm}
\bibliography{references}

\end{document}